\documentstyle[12pt]{article}

\def\hybrid{\topmargin -20pt  \oddsidemargin 0pt
      \headheight 0pt   \headsep 0pt
      \textwidth 6.25in % A4 paper
      \textheight 9.5in % A4 paper
      \marginparwidth .875in
      \parskip 5pt plus 1pt   \jot = 1.5ex}

\hybrid

\begin{document}
%\titlepage
\def\x{\times}
\def\beq{\begin{equation}}
\def\eeq{\end{equation}}
\def\beqa{\begin{eqnarray}}
\def\eeqa{\end{eqnarray}}

\sloppy
\newcommand{\be}{\begin{equation}}
\newcommand{\eq}{\end{equation}}
\newcommand{\ov}{\overline}
\newcommand{\un}{\underline}
\newcommand{\p}{\partial}
\newcommand{\la}{\langle}
\newcommand{\ra}{\rangle}
\newcommand{\bl}{\boldmath}
\newcommand{\ds}{\displaystyle}
\newcommand{\nl}{\newline}
\newcommand{\th}{\theta}

%\textwidth14.5cm
%\textheight23.0cm
%\oddsidemargin0.5cm
%\topmargin-1.4cm

%\addtocounter{section}{1}

\renewcommand{\thesection}{\arabic{section}}
\renewcommand{\theequation}{\thesection.\arabic{equation}}

%\setcounter{section}{1}
%\addtocounter{section}{1}
\parindent0em

%\newcommand{\resetcounter}{\setcounter{equation}{0}}     
% set counter to zero

\begin{titlepage}
\begin{center}
\hfill HUB-EP-97/31\\
\hfill {\tt hep-th/9705174}\\

\vskip .7in

{\bf  $N=1$ DUAL STRING PAIRS AND THEIR MASSLESS SPECTRA}

\vskip .3in

Bj\"orn Andreas, Gottfried Curio and Dieter L\"ust\footnote{email:
andreas@qft3.physik.hu-berlin.de, curio@qft2.physik.hu-berlin.de, 
luest@qft1.physik.hu-berlin.de}
\\
\vskip 1.2cm

{\em Humboldt-Universit\"at zu Berlin,
Institut f\"ur Physik, 
D-10115 Berlin, Germany}

\vskip .1in

\end{center}

\vskip .2in

%\begin{center} {\bf ABSTRACT } \end{center}
\begin{quotation}\noindent

We construct two chains of fourdimensional
$F$-theory/heterotic dual string pairs with
$N=1$ supersymmetry. On the $F$-theory side
as well as on the heterotic side the geometry of the 
involved manifolds relies on del Pezzo surfaces. 
We match the massless spectra
by using, for one chain of models, an index formula
to count the heterotic bundle moduli and 
determine the dual $F$-theory spectra from the Hodge numbers
of the fourfolds $X^4$ and of the type IIB base spaces.

\end{quotation}
\end{titlepage}
\vfill
\eject

\newpage

\section{Introduction}

Various type of duality symmetries in compactified string vacua
were established during recent years; moreover
several transitions among different vacua were explored.
The unification of possibly all superstring 
theories is expected to be
realized by a common underlying framework, often called $M$-theory.
For example, strong/weak coupling duality symmetries 
among type II compactifications on Calabi-Yau threefolds
and heterotic vacua on $K3\times T^2$
with $N=2$ supersymmetry in four dimensions \cite{KV}
were successfully tested in a broad class of models.
Using these $N=2$ string duality symmetries results
about nonperturbative effects in string theory and field theory
can be derived, like the computation of the nonperturbative
heterotic $N=2$ prepotential or nonperturbative effects in $N=2$
field theory like Seiberg/Witten \cite{SW}.
Of even greater phenomenological interest is the investigation
of string duality symmetries in fourdimensional string vacua with
$N=1$ space-time supersymmetry. One hopes in this way to get
important nonperturbative information, like the computation of
nonperturbative $N=1$ superpotentials \cite{W} 
and supersymmetry breaking, or the
stringy reformulation of many effects in $N=1$ field theory.
In addition, it is a very important question, how 
transitions among $N=1$ string vacua,
possibly with a different number of chiral multiplets, take place.

String dual pairs with $N=1$ supersymmetry in four dimensions 
\cite{KatzV,BJPSV},[\ref{wit},\ref{BJPS}]
are
provided by comparing heterotic string vacua on  Calabi-Yau threefolds 
$Z$,
supplemented by the specification of a particular choice of the
heterotic gauge bundle $V\rightarrow Z$, 
with particular fourdimensional type IIB
superstring vacua, which can be formulated as $F$ theory
compactifications \cite{V,MV} on elliptic Calabi-Yau fourfolds $X^4$.
To be more specific, the elliptic fibration of the Calabi-Yau fourfold 
has a (complex) threedimensional base manifold $B^3$,
i.e. $X^4\rightarrow B^3$, which is 
the (non Calabi-Yau) type IIB  
compactification space from ten to four dimensions.
Moreover one supposes that
$B^3$ is rationally ruled, i.e. there exists a fibration
$B^3\rightarrow S$ with $P^1$ fibers, because one has to assume 
the four-fold $X^4$ to be a $K3$ fibration
over the two dimensional surface $S$, i.e.
$Z\rightarrow S$.  
Then, using the eightdimensional
duality among $F$-theory on an elliptic $K3$ and the heterotic string
on $T^2$, one derives by an adiabatic argument that the heterotic
Calabi-Yau is an elliptic fibration
over the same surface $S$.
In this paper we will explicitly construct $N=1$ dual $F$-theory/heterotic
string pairs with $S$ being the socalled del Pezzo surfaces
$dP_k$, the $P^2$'s blown up in $k$ points. Then, in our class
of models the three-dimensional bases, $B^3_{n,k}$, are characterized
by two parameters $k$ and $n$, where $n$ encodes the fibration
structure $B^3_{n,k}\rightarrow dP_k$.
Specifically we will discuss the cases $n=0$, $k=0,\dots ,6$, i.e.
$B^3_{0,k}=P^1\times dP_k$, and also the cases $n=0,2,4,6,12$, $k=9$, 
where the 9 points are the intersection of two cubics and 
the $B_{n,9}$ are a fibre product
$dP_9\times_{P^1} F_{n/2}$ with common base $P^1$ and $F_{n/2}$ the 
rational ruled Hirzebruch surfaces.
We will show that the massless spectra match on both sides.

The considered class of models is the direct generalization
of the $F$-theory/heterotic $N=1$ dual string pair with $n=0$, $k=9$ 
(again the 9 points in the mentioned special position),
which we investigated in \cite{CL}, which was a ${\bf Z}_2$ modding
of a $N=2$ model described as $F$-theory on $T^2\times CY^3$ (also
considered there was the ${\bf Z}_2$ modding of $K3\times K3$). In this model
we also considered a duality check for the superpotential.

The first chain of models we will discuss here consists then in
heterotic Calabi-Yau's elliptic over $dP_k$ ($k=0,\dots ,3$) with
a {\it general} $E_8\times E_8$ vector bundle resp. $F$-theory on smooth
Calabi-Yau fourfolds
$X^4_k$ elliptic over $dP_k \times P^1$. The heterotic
Calabi-Yau of $\chi\neq 0$ varies with $k$ and one can also see
the transition induced from the blow-ups in $P^2$ among the
fourfolds on the $F$-theory side (for transitions among $N=1$ vacua cf. also
[\ref{KS}]).

Then we go on to construct a second chain of fourdimensional
$F$-theory/heterotic dual string pairs with
$N=1$ supersymmetry by $Z_2$-modding of corresponding dual pairs 
with $N=2$ supersymmetry.
The resulting Calabi-Yau four-folds $X^4_n$ are $K3$-fibrations over the
del Pezzo surface $dP_9$ (with points in special position).
On the heterotic side the dual models are obtained by compactification
on a known Voisin-Borcea Calabi-Yau three-fold with
Hodge numbers $h^{(1,1)}=h^{(2,1)}=19$, where,
similarly as in the underlying $N=2$ models, the 
heterotic gauge bundles over this space are characterized by turning on
$(6+{n\over 2},6-{n\over 2})$ instantons of $E_8\times E_8$.
We work out the Higgsing chains of the 
gauge groups together with their massless
matter content (for example the numbers of 
chiral multiplets in the ${\bf 27}+{\bf\overline{27}}$ representation
of $E_6$) for each model and 
show that the  heterotic spectra
of our models match the
dual $F$-theory spectra, as computed from the Hodge numbers
of the four-folds $X^4_n$ and of the type IIB base spaces.

Note that in contrast to this second class of models the first chain
(varying $k$, $n=0$) consists of {\it genuine} $N=1$ models.
Moreover, within the first chain the heterotic Calabi-Yau spaces have
non-vanishing Euler numbers, potentially leading to theories
with chiral spectra with respect to  non-Abelian gauge groups which show
up at certain points in the moduli spaces. Hence the transitions in $k$
might connect $N=1$ models with different numbers of chiral fermions.

In section 2 we consider $F$-theory on smooth fourfolds $X^4_k$ elliptic
over $dP_k \times P^1$ and propose an identification of the spectrum
from the Hodge numbers of a general 
$X^4$ as well as of its threefold base $B^3$.\\
In section 3 we study a class of Calabi-Yau threefolds elliptic
over $dP_k$ and derive an index formula for the number of moduli of
a general $E_8 \times E_8$ bundle.\\
In section 4 we consider the ${\bf Z}_2$ modding of the $N=2$ models
described by $F$-theory on $T^2 \times X^3_n$, where $X^3_n$ are the well
known threefold Calabi-Yau over $F_n$.\\
In section 5 we make the corresponding modding in the dual heterotic $N=2$
model on $T^2 \times K3$ with instanton embedding $(12+n,12-n)$.

\section{$F$-theory on ${\bf X^4_k}\rightarrow {\bf P}^1\times {\bf dP}_k$}
\setcounter{equation}{0}

We consider F-theory on a smooth elliptically fibered Calabi-Yau fourfold 
$X^4$ with base $B^{3}_k={\bf P}^1\times {dP}_k$ which can be represented 
by a smooth Weierstrass model over $B^{3}_k$ if the anti-canonical
line bundle $-K_{B}$ over $B^{3}_k$ is very ample [\ref{gras}].
The Weierstrass model is 
described by the homogeneous equation $y^{2}z=x^{3}+g_{2}xz^{2}+g_{3}z^{3}$
in a ${\bf P}^2$ bundle $W\rightarrow B^{3}_k$. The ${\bf P}^2$ bundle 
is the projectivization of a vector bundle $K_{B}^{-2}\oplus
K_{B}^{-3}\oplus{\cal O}$ over $B^{3}_k$. 
Furthermore we can
think of $x$,$y$ and $z$ as homogeneous 
coordinates on the ${\bf P}^2$ fibres, i.e.
they are sections of ${\cal O}(1)\otimes K_{B}^{-2}$, ${\cal O}(1)
\otimes{K}_{B}^{-3}$ and ${\cal O}(1)$ over $W$ respectively; 
$g_{2}$ and $g_{3}$ are sections of $H^{0}(B^{3}_{k},K_{B}^{-4})$ and  
$H^{0}(B^{3}_{k}, K_{B}^{-6})$ [\ref{fri}].\\
Most of the 84 known Fano threefolds have a very ample $-K_{B}$ 
[\ref{gras}]. 18 of them are toric threefolds ${\cal F}_n$ (
$1\le n \le 18$) and are completely classified [\ref{its1}, \ref{its2},
\ref{MM}, \ref{MM2}]. They were recently studied in the context of 
Calabi-Yau fourfold compactifications over Fano threefolds [\ref{klem}, 
\ref{mohri}]. In particular in [\ref{gras}] it was shown that for 
$k=0,...,6$ over $B^{3}_k$ there exists a smooth Weierstrass model 
(having a section); $B^{3}_k$ with $k=0,1,2,3$ 
correspond to the toric Fano threefolds ${\cal F}_n$ with $n=2, 9, 13, 17$.
The corresponding fourfolds $X^4$ have a $K3$ fibration over ${dP}_k$ 
cf. [\ref{mohri},\ref{mayr}].
First we can determine the Hodge numbers of $B^{3}_k$.
The number of K\"ahler and complex structure parameters of $B^{3}_k$ are 
$h^{11}(B^{3}_k)=2+k$ (where the 2 is coming from the line in ${\bf P}^2$ and 
the class of the ${\bf P}^1)$ and $h^{21}(B^{3}_k)=0$.
For $k=0,...,6$ we can compute the Euler number of X in terms of topological 
data of the base according to [\ref{sethi}]

\begin{eqnarray}
\chi=12\int_{B^{3}_k}c_1c_2+360\int_{B^{3}_k}c_{1}^3
\end{eqnarray}

where the $c_{i}$ refer to $B^{3}_k$ (the 360 is related to the
Coxeter number of $E_8$ which is associated with the elliptic fiber
type [\ref{klem}]).\\

One finds $\int_{B^{3}_k}c_1c_2=24$ and 
$\int_{B^{3}_k}c_{1}^3= 3c_{1}({\bf P}^1)c_{1}(dP_k)^2= 6(9-k)$ which
leads to 

\begin{eqnarray}
\chi=288+2160(9-k).
\end{eqnarray}

In the following we restrict to
$k=0,1,2,3$ where one has $h^{21}=0$, [\ref{klem},\ref{mohri}],
$h^{11}(X^4_k)=3+k$, $h^{31}(X^4_k)=\frac{\chi}{6}-8-(3+k)=
28+361(9-k)$ [\ref{klem},\ref{mohri},\ref{sethi}].

Next let us compute for a {\it general} model 
from the Hodge numbers of $X^4$ the spectrum
of massless $N=1$ superfields in the $F$-theory compactification.
Just as in the six-dimensional case (see Ch. 4.1), 
also the Hodge numbers of the
type IIB bases $B^3$, i.e.
the details of the elliptic fibrations, 
will enter the numbers of massless fields.
So consider first the compactification of the type IIB string from ten
to four dimensions on the spaces $B^3$.
Abelian $U(1)$ $N=1$ vector multiplets
arise from the dimensional reduction of the four-form antisymmetric
Ramond-Ramond tensor field $A_{MNPQ}$ in ten dimensions; therefore we expect
that the rank of the four-dimensional gauge group, $r(V)$, gets contributions
from the $(2,1)$-forms of $B^3$ such that $h^{(2,1)}(B^3)$ contributes
to $r(V)$. Chiral (respectively anti-chiral) $N=1$ multiplets,
which are uncharged under the gauge group, arise from from
$A_{MNPQ}$ with two internal Lorentz indices as well from the
two two-form fields $A^{1,2}_{MN}$ (Ramond-Ramond and NS-NS) with zero or
two internal Lorentz indices. Therefore we expect that the number of
singlet chiral fields, $C$, receives contributions from $h^{(1,1)}(B^3)$.
On the other hand we can study the $F$-theory spectrum in three dimensions
upon futher compactification on a circle $S^1$. This is equivalent to
the compactification of 11-dimensional supergravity on the same $X^4$.
(Equivalently we could also consider the compactifications of the
IIA superstring on $X^4$ to two dimensions.) So in three dimensions the 
11-dimensional three-form field $A_{MNP}$ contributes
$h^{(1,1)}(X^4)$ to $r(V)$ and $h^{(2,1)}$ to $C$. In addition,
the complex structure deformations of the 11-dimensional metric contributes 
$h^{(3,1)}(X^4)$ chiral fields. (These fields arise in analogy to the
$h^{(2,1)}$ complex
scalars which describe the complex structure of the metric when compactifying
on a Calabi-Yau three-fold from ten to four dimensions.)
Since, however, vector and chiral fields are
equivalent in three dimensions by Poincare duality, this implies that the
sum $r(V)+C$ must be independent from the Hodge numbers of the type IIB bases 
$B^3$. Therefore, just in analogy to the
six-dimensional $F$-theory compactifications, the following formulas for
the spectrum of the four-dimensional $F$-theory models on Calabi-Yau
four-folds are expected \cite{CL},[\ref{mohri}]: namely for the rank of
the $N=1$ gauge group we derive
\beqa
r(V)=h^{(1,1)}(X^4)-h^{(1,1)}(B^3)-1+h^{(2,1)}(B^3),\label{spectrumv} 
\eeqa
and for the number $C$ of $N=1$ neutral chiral 
(resp. anti-chiral) multiplets we get
\beqa
C&=&h^{(1,1)}(B^3)-1+h^{(2,1)}(X^4)-h^{(2,1)}(B^3)+h^{(3,1)}(X^4)
\nonumber\\
 &=&h^{(1,1)}(X^4)-2+h^{(2,1)}(X^4)+h^{(3,1)}(X^4)-r(V)\nonumber\\
 &=&
 \frac{\chi}{6}-10+2h^{(2,1)}(X^4)-r(V).
 \label{spectrumc}
\eeqa
Note that in this formula we did not count the chiral field which 
corresponds to the dual heterotic dilaton.

Now let us apply these formulae to the fourfolds $X^4_k$.
For the rank of the N=1 gauge group we derive 

\begin{eqnarray}
r(V)&=& h^{11}(X^4_k)-h^{11}(B^3_k)-1+h^{21}(B^3_k)\nonumber\\
    &=& 0
\end{eqnarray}

and for the number of N=1 neutral chiral (resp. anti-chiral)
multiplets $C_{F}$ we get

\begin{eqnarray}
C_{F}&=&h^{11}(B^3_k)-1+h^{21}(X^4_k)-h^{21}(B^3_k)+h^{31}(X^4_k)\nonumber\\
     &=&38+360(9-k)
\end{eqnarray}

\begin{center}
\begin{tabular}{|c||c|c|c|c|c|} \hline
k & $\chi$ & $h^{31}$ & $h^{22}$ & $h^{11}$ & $C_{F}$ 
\\  \hline 
0 & 19728 & 3277 & 13164 &  3  & 3278 \\  \hline
1 & 17568 & 2916 & 11724 &  4  & 2918 \\  \hline
2 & 15408 & 2555 & 10284 &  5  & 2558 \\  \hline 
3 & 13248 & 2194 &  8844 &  6  & 2198 \\  \hline
4 & 11088 & 1833 &  7404 &  7  & 1838 \\  \hline
5 &  8928 & 1472 &  5964 &  8  & 1478 \\  \hline
6 &  6768 & 1111 &  4524 &  9  & 1118 \\  \hline

\end{tabular}
\end{center}

In this table we have made for the cases $k=4,5,6$ the assumption 
$h^{21}=0$. With this assumption 
the matching we will present in this paper goes through also in
these cases.\\ 

Before we come to the heterotic side let us make some comments.
A consistent $F$- theory compactification on a Calabi-Yau fourfold
requires the presence of $\frac{\chi}{24}$ threebranes in the 
vacuum [\ref{sethi}]. Under duality with the heterotic string, the 
threebranes turn into fivebranes wrapping the elliptic fibers and 
in [\ref{wit}] it was shown that one can understand
the appeareance of the fivebranes by using the non-perturbative
anomaly cancellation condition with fivebranes. The non-perturbative
anomaly cancellation condition is given by [\ref{wit}]

\begin{eqnarray}
\lambda(V_{1})+\lambda(V_{2})+[W]=c_2(TZ)
\end{eqnarray}

here $TZ$ is the tangent bundle of the Calabi-Yau threefold $Z$ on which
the heterotic string is compactified, $V_{1}\times V_{2}$ is a 
$E_{8}\times E_{8}$ bundle with $\lambda(V)$ denoting its fundamental
characteristic class and $[W]$ is the cohomology class of the fivebranes
where $[W]=h[F]=\pi^{*}([p])$ and $[F]$ is the class in $H^{4}$ of the 
fiber of the 
elliptic fibration and h is the number of
fivebranes on the heterotic string side. In order to show that $h=\frac{\chi}{24}$ one 
has to express the number of threebranes in terms of topological data
defined on $B^{2}$. These was done by Friedman, Morgan and Witten 
[\ref{wit}]. They considered the $F$-theory base $B^3$ as a ${\bf P}^1$ bundle
over $B^2$ being the projectivization of a vector
bundle ${\cal O}\oplus {\cal T}$ with ${\cal O}$ and ${\cal T}$ being 
line bundles over $B^2$ and $r=c_{1}({\cal O}(1))$,$t=c_{1}({\cal T})$
(note that $t$ plays the role of the $n$ of ${F_{n}}$ in the 
N=2 string dualities in six dimensions between the heterotic string 
on $K3$ with ($12+n, 12-n$) instantons embedded in each $E_{8}$ and 
$F$ theory on elliptic fibered Calabi-Yau threefold over ${F_{n}}$).
In our case we have $t=0$. For the number of threebranes 
one has [\ref{wit}] :

\begin{eqnarray}
\frac{\chi}{24}=\int_{B^{2}}(c_{2}+91c_{1}^2+30t^2).
\end{eqnarray}
 
which simplifies in our case using $t=0$. Finally for our base $B^{2}$
we can write:

\begin{eqnarray}
\frac{\chi}{24}=822-90k.
\end{eqnarray}

 From the last expression we learn that between each blow up there is a 
threebrane difference of 90. 
Note that Sethi, Vafa and Witten [\ref{sethi}] had a brane difference of 120
as they blow up in a threefold whereas we do this in $B^2$.

\section{Heterotic string on ${\bf Z}\rightarrow {\bf dP}_k$}

To compactify the heterotic string on elliptically fibered Calabi Yau
threefold we have in addition to specify a vector bundle V with fixed
second Chern class. Since we had zero for the rank $r(V)$ of the N=1 
gauge group on the $F$ theory 
side we have to switch on a $E_{8}\times E_{8}$ bundle that breaks the
gauge group completely.\\ 
Now let $Z$ be a nonsingular elliptically fibered Calabi-Yau threefold over
${dP}_k$ with a section. Recall the Picard group of ${dP}_k$
is $Pic$ ${dP}_k={\bf Z}{\ell}\oplus{\bf Z}{\ell}_i$ where ${\ell}$ denotes 
the class of the line in  ${\bf P}^2$ and ${\ell}_i$, $i=1,...,k$ are the 
classes of the blown up points. The intersection form is 
defined by [\ref{Manin}]

\begin{eqnarray}
{\ell}^2=1, \ \ \ {\ell}_{i}\cdot{\ell}=0 , 
\ \ \  {\ell}_{i}\cdot {\ell}_{j}=-\delta_{ij}, \ \
\end{eqnarray} 

and the canonical class of ${dP}_k$ is 

\begin{eqnarray}
{\cal K}_{\cal B}=-3{\ell}+\sum_{i=1}^{k}{\ell}_i .
\end{eqnarray}

Assuming again that we are in the case of a 
general\footnote {i.e. having only 
one section [\ref{gras}], so a typical counterexample would be the
$CY^{19,19}=B_{9}\times _{P^1}B_{9}$ with the 9 points being the intersection
of two cubics} smooth Weierstrass model one has $h^{11}(Z)=2+k$. So the 
fourth homology of $Z$ is generated by the following divisor
classes: $D_0=\pi^*{\ell}$, $D_i=\pi^*{\ell}_i$ and $S=\sigma({\cal B}_k)$.
The intersection form on $Z$ is then given by

\begin{eqnarray}
S^3=9-k, \ \ D_{0}^2S=1, \ \ D_{i}^2S=-1, \ \ 
                              D_{i}S^2=-1, \ \ D_{0}S^2=-3,  
\end{eqnarray}   

all other triple intersections are equal to zero.\\
The canonical bundle for $Z\rightarrow B$ is given by \cite{MV}

\begin{eqnarray}
{\cal K}_{Z}=\pi^{*}({\cal K}_{B}+\sum a_{i}[\Sigma_{i}])  
\end{eqnarray} 

where $a_{i}$ are determined by the type of singular fiber and $\Sigma_{i}$
is the component of the locus within the base on wich the elliptic curve 
degenerates. In order to get ${\cal K}_{Z}$ trivial one requires 
${\cal K}_{B}=-\sum a_{i}[\Sigma_{i}]$. \\
Since we are interested in a smooth elliptic fibration we have to check
that the elliptic fibration does not degenerate more worse than with $I_{1}$
singular 
fiber over codimension one in the base which then admits a smooth
Weierstrass model. That this indeed happens in our case of del Pezzo base
can be proved by similar methods as for the $F_{n}$ case in \cite{MV}.
As $a_{i}=-1$ we have a smooth elliptic fibration.\\
The Euler number of $Z$ is given by (assuming $E_8$ elliptic fiber type, cf.
[\ref{klem}]) 

\begin{eqnarray}
\chi(Z)=-60 \int c_{1}^2(B)
\end{eqnarray}

and\footnote{Note that the transition by 29 in $h^{21}(Z)$ in going from
$k$ to $k+1$ has a well known interpretation if one uses these Calabi-Yau
threefolds as $F$-theory compactification spaces cf. \cite{CF}}

\begin{eqnarray}
h^{11}(Z)&=&2+k=11-(9-k) \\
h^{21}(Z)&=&(272-29k)=11+29(9-k).
\end{eqnarray}

In order to compare the spectrum of the heterotic side with the 
$F$ theory side we have still to determine the number of moduli coming 
from the $E_{8}\times E_{8}$ bundle over $Z$. 
Therefore we use the 
${\bf Z}_{2}$ charactervalued index theorem [\ref{wit}]
using the fact that a elliptic manifold with a section has a ${\bf Z}_{2}$
symmetry generated by an involution $\tau$ that leaves the section 
invariant and acts as -1 on each fiber or in terms of the Weierstrass 
model as $y\rightarrow -y$. As $h^{21}(X^4)=0$ the index gives the 
number of bundle moduli [\ref{wit}] (we assume no 4-flux has been turned
on (which is not a free decision in general [\ref{flux}]))

\begin{eqnarray}
I=rk-\sum_{i}\int_{U_{i}}\lambda(V)
\end{eqnarray} 

The $U_{i}$ are the components of the fixed point set. $U_{1}$ is
given by $x=z=0, y=1$ and is the section of $Z\rightarrow B$; the second
component, $U_{2}$, is given by $y=0$ and is a triple cover of $B$ given
by the trisection $0=4x^3-g_{2}xz^2-g_{3}z^3$ which is defined in a 
${\bf P}^1$ bundle $W^\prime$ over $B$ where $x$, $z$ are sections of 
${\cal O}(1)\otimes K_{B}^{-2}$ and ${\cal O}(1)$ respectively with 
$c_{1}({\cal O}(1))=r$ and $c_{1}({\cal O}(1)\otimes K_{B}^{-2})=r+2c_{1}$. 
Furthermore $\lambda(V)$ is the
fundamental characteristic class of the $E_{8}\times E_{8}$ bundle.\\
Now one has the identity 
$\int_{U_{i}} \lambda (V)= \int_{B} \lambda (V)\mid_{U_{1}}+ 3\int_{B} 
\lambda (V)\mid_{U_{2}}$ where the 3 appears since $U_{2}$ is a triple
cover of $B$. One has

\begin{eqnarray}
\lambda(V_{1})\mid_{U_{i}}+
\lambda(V_{2})\mid_{U_{i}}+[W]\mid_{U_{i}}=c_2(TZ\mid_{U_{i}})
\end{eqnarray}

Now the restriction of the tangent bundle to $U_{i}$ can be derived by the 
exact Euler sequence 
$0\rightarrow{\cal T}U_{i}\rightarrow{TZ\mid_{U_{i}}}\rightarrow
{\cal N}_{Z}U_{i}\rightarrow 0$ where ${\cal N}_{Z}{U_{i}}$ 
is the normal bundle to $U_{i}$ in $Z$.
As $U_{1}$ is the section of $Z\rightarrow
{dP}_{k}$ we can identify $U_{1}$ with the base and express the total
Chern class of $TZ\mid_{U_{1}}$ in terms of $c_{i}=\pi^{*}(c_{i}({dP}_k)$.
We find

\begin{eqnarray}
c_{1}(TZ\mid_{U_{1}})&=&c_{1}+c_{1}({\cal N}_{Z}{U_{1}})\\
c_{2}(TZ\mid_{U_{1}})&=&c_{2}+c_{1}({\cal N}_{Z}{U_{1}})c_{1}({dP}_k)
\end{eqnarray} 

and using the Calabi-Yau condition $c_1=-c_1({\cal N}_{Z}{U_{1}})$ we find
for the restricted second Chern class:

\begin{eqnarray}
c_{2}(TZ\mid_{U_{1}})=c_{2}-c_{1}^2.
\end{eqnarray}

Now we have to determine $c_{2}(TZ\mid_{U_{2}})$. Therefore recall that $U_{2}$
was given by $y=0$ defining a triple cover of ${dP}_{k}$ in a ${\bf P}^1$
bundle $W^{\prime}\rightarrow {dP}_{k}$. To determine $c_{2}(TZ\mid_{U_{2}})$  
we use techniques from [\ref{sethi}]. The cohomology ring of $W^{\prime}$ is
generated over the cohomology ring of ${dP}_{k}$ by the element 
$r=c_1({\cal O}(1))$ with the relation $r(r+2c_{1})=0$ which expresses the fact
that $x$ and $z$ have no common zero. Since $U_{2}$ is defined by the 
vanishing of a
section of ${\cal O}(1)^{3}\otimes K_{dP_{k}}^{-6}$ which is a line bundle 
over $W^{\prime}$ with $c_{1}({\cal O}(1)^{3}\otimes K_{dP_{k}}^{-6})=
3r+6c_{1}$. Any cohomology class on $U_{2}$ that can be extended over $W^{\prime}$
can be integrated over $U_{2}$ by multiplying it with the cohomology class of 
$U_{2}$ in $H^2(Z)$ which is $3r+2c_{1}$ cf. [\ref{wit}], 
i.e. multiplication by $3r+2c_{1}$
can be understood as restriction from $W^{\prime}$ to $U_{2}$. The relation 
$r(r+2c_{1})$ can then be simplified to $r=0$. Now we are able to compute the
total Chern class of the projective space bundle $W^{\prime}$ given by the 
adjunction formula $c(W^{\prime})=c({dP}_{k})(1+r)(1+r+2c_{1})$. 
Thus the total Chern class
of the tangent bundle of $U_{2}$ is obtained by dividing by $(1+3r+6c_{1})$
and using $r=0$

\begin{eqnarray}
c({\cal T}U_{2})=c({dP}_{k})\frac{(1+2c_{1})}{(1+6c_{1})}.
\end{eqnarray}

 From the total Chern class we derive 

\begin{eqnarray}
c_{1}(TZ\mid_{U_{2}})&=&-3c_{1}+c_{1}({\cal N}_Z{U_{2}})\\
c_{2}(TZ\mid_{U_{2}})&=& 20c_{1}^{2}+c_{2}-3c_{1}+c_{1}({\cal N}_Z{U_{2}})
\end{eqnarray}

and with the Calabi-Yau condition $3c_{1}=c_{1}({\cal N}_Z U_{2})$ we
get 

\begin{eqnarray}
c_{2}(TZ\mid_{U_{2}})=c_{2}+11c_{1}^2.
\end{eqnarray}

Taking into account the restricted five brane correction term $h=\int_{B}
c_{2}+91c_1^2+30t^2$
we find for the index a formula first derived by E. Witten [\ref{un}]\footnote
{We thank S. Kachru for pointing out this reference to us.}

\begin{eqnarray}
I=16+332\int_{B}c_{1}^2(B)+120\int_{B}t^2
\end{eqnarray}  

Remember that in our case the last term vanishes and we find for the number
of bundle moduli

\begin{eqnarray}
I=16+332(9-k)
\end{eqnarray}

For the number of $N=1$ neutral chiral (resp. antichiral) multiplets
$C_{{\rm het}}$ we find

\begin{eqnarray}
C_{{\rm het}}&=& h^{21}(Z)+h^{11}(Z)+I\nonumber\\
   &=& 38+360(9-k)
\end{eqnarray}

which agrees with the number of chiral multiplets $C_F$ 
eq.(2.6) on the $F$-theory side.

\begin{center}
\begin{tabular}{|c||c|c|c|c|} \hline
k & $\chi$ & $h^{21}$ & $h^{11}$ & $I$\\  \hline 
0 & -540 & 272 &  2  & 3004    \\  \hline
1 & -480 & 243 &  3  & 2672    \\  \hline
2 & -420 & 214 &  4  & 2340    \\  \hline 
3 & -360 & 185 &  5  & 2008    \\  \hline
4 & -300 & 156 &  6  & 1676    \\  \hline
5 & -240 & 127 &  7  & 1344    \\  \hline
6 & -180 &  98 &  8  & 1012    \\  \hline
\end{tabular}
\end{center}

\section{$F$-theory on the $K3$-fibred four-folds $X^4_n$}
\setcounter{equation}{0}

\subsection{The $N=2$ models: $F$-theory on $X^3_n(\times T^2)$}

We will start constructing the four-folds $X^4_n$
by first considering $F$-theory compactified
to six dimensions on an elliptic Calabi-Yau threefold $X^3$, and then
further to four dimensions on a two torus $T^2$, i.e. the
total space is given by $X^3\times T^2$.
This leads to $N=2$ supersymmetry in four dimensions.
As explained in \cite{V}, this four-dimensional $F$-theory compactification
is equivalent to the type IIA string compactified on the same
Calabi-Yau $X^3$. 
 
To be more specific let us discuss the cases where the Calabi-Yau
threefolds,
which we call $X^3_n$, are elliptic fibrations over 
the rational ruled Hirzebruch
surfaces $F_n$ \cite{MV}; 
the Hirzebruch surfaces $F_n$
with complex coordinates $z_1$ and
$z_2$ are $P^1_{z_2}$ fibrations over $P^1_{z_1}$.
The corresponding type IIB base spaces are given
by the Hirzebruch surfaces $F_n$ in six dimensions.
In the following it will become very useful to describe the elliptically
fibred Calabi-Yau spaces $X^3_n$ in the Weierstrass form \cite{MV}:
\beqa
X^3_n:\quad 
y^2=x^3+\sum_{k=-4}^4f_{8-nk}(z_1)z_2^{4-k}x+\sum_{k=-6}^6g_{12-nk}(z_1)
z_2^{6-k}.\label{weier}
\eeqa
Here $f_{8-nk}(z_1)$, $g_{12-nk}(z_1)$ are polynomials of degree
$8-nk$, $12-nk$ respectively, where the polynomials with negative degrees
are identically set to zero. From this equation we see that the Calabi-Yau
threefolds
$X^3_n$ are $K3$ fibrations over $P^1_{z_1}$ with coordinate $z_1$; the 
$K3$ fibres themselves are elliptic fibrations over the $P^1_{z_2}$ with
coordinate $z_2$.

The Hodge numbers $h^{(2,1)}(X^3_n)$, 
which count the number of complex structure
deformations of $X^3_n$, are the given by the the number of parameters
of the curve (\ref{weier}) minus the number of possible reparametrizations,
which are given by 7 for $n=0,2$ and by $n+6$ for $n>2$.
On the other hand, the Hodge numbers $h^{(1,1)}(X^3_n)$, which count the number
of  K\"ahler parameters of $X^3_n$, are determined by the Picard number
$\rho$ of the $K3$-fibre of $X^3_n$ as
\beqa
h^{(1,1)}(X^3_n)=1+\rho.\label{h11x3}
\eeqa
Let us list the Hodge numbers of the $X^3_n$ for those cases which are
relevant for our following discussion:
\hspace{0.2cm}

\beqa
%\begin{center}
\begin{tabular}{|c||c|c|} \hline
  $n$     & $h^{(1,1)}(X^3_n)$ & $h^{(2,1)}(X^3_n)$ \\ \hline\hline
0  &   3 & 243   \\ \hline
2 &   3 & 243    \\ \hline
4 &    7 & 271 \\ \hline
6 & 9 & 321 \\ \hline
12 & 11 & 491 \\ \hline
\end{tabular}
%\end{center}
\label{tab1}
\eeqa
%\hspace{0.5cm}

Let us also recall \cite{V,MV} how the Hodge numbers 
of $X^3_n$ determine the spectrum
of the $F$-theory compactifications. In six dimensions
the number of tensor multiplets $T$ is given by the number of
K\"ahler deformations of the (compex) 
two-dimensional type IIB base $B^2$ except
for the overall volume of $B^2$:
\beqa
T=h^{(1,1)}(B^2)-1.\label{notensor}
\eeqa
These tensor fields become Abelian $N=2$ vector fields upon further
$T^2$ compactification to four dimensions. Since the four-dimensional $F$-theory
is equivalent to the type IIA string on $X^3_n$ it follows that the
number of four-dimensional Abelian vector fields in the Coulomb phase
is given by $T+r(V)+2=h^{(1,1)}(X^3_n)$, where $r(V)$ is the rank of the
six-dimensional gauge group, and the additonal two vector fields
arise from the toroidal compactification.
This then leads to the following equation for $r(V)$:
\beqa
r(V)=h^{(1,1)}(X^3_n)-h^{(1,1)}(B^2)-1.\label{novector}
\eeqa
Finally, the number of hypermultiplets $H$, which are neutral under the Abelian
gauge group, is given in four as well as in six dimensions
by the number of complex deformations of $X^3_n$ 
plus the freedom in varying the the size of the base $B^2$:
\beqa
H=h^{(2,1)}(X^3_n)+1.\label{nohyper}
\eeqa
For the cases we are interested in, namely $B^2=F_n$, $h^{(1,1)}(F_n)$ is
universally given by $h^{(1,1)}(F_n)=2$. Therefore one immediately gets
that $T=1$, which corresponds to the universal dilaton tensor multiplet
in six dimensions, and 
\beqa
r(V)=\rho -2=h^{(1,1)}(X^3_n)-3.\label{novector1}
\eeqa

At special loci in the moduli spaces of the hypermultiplets one
obtains enhanced non-Abelian gauge symmetries. These loci are determined 
by the 
singularities of the curve (\ref{weier}) and were analyzed in detail in
\cite{BIKMSV}. These $F$-theory
singularities correspond to the perturbative gauge symmetry enhancement
in the dual heterotic models.

\subsection{The $N=1$ models: $F$-theory on $X^4_n=(X^3_n\times T^2)/Z_2$}

Now we will construct  from the $N=2$ $F$-theories on
$X^3_n\times T^2$ the corresponding $N=1$ models on Calabi-Yau four-folds 
$X^4_n$ by a $Z_2$ modding procedure, i.e.
\beqa
X^4_n={X^3_n\times T^2\over Z_2}.\label{modding}
\eeqa
%The precise definition of the $Z_2$ modding follows from the corresponding
%operation on the heterotic side (see
%next chapter), as we have already constructed
%the four-fold $X^4_0$ by this procedure
%in \cite{CL}.
First, the $Z_2$ modding acts as quadratic redefinition
on the coordinate $z_1$, the coordinate
of the base $P^1_{z_1}$ of the $K3$-fibred space $X^3_n$,
i.e. the operation is $z_1\rightarrow -z_1$. This means that
the modding is induced from the quadratic base map
$z_1\rightarrow \tilde z_1:=z_1^2$ with the two branch points 0 and $\infty$.
So the degrees of the corresponding polynomials $f(z_1)$ and $g(z_1)$
in eq.(\ref{weier}) are reduced by half (i.e. the moddable cases are the
ones where only even degrees occur).
So instead of the Calabi-Yau threefolds $X^3_n$ we are now dealing
with the non-Calabi-Yau threefolds ${\cal B}^3_n=X^3_n/Z_2$ which can
be written in Weierstrass form as follows:
\beqa
{\cal B}^3_n:\quad y^2=x^3+\sum_{k=-4}^4f_{4-{nk\over 2}}
(z_1)z_2^{4-k}x+\sum_{k=-6}^6g_{6-{nk\over 6}}(z_1)
z_2^{6-k}.\label{weierb}
\eeqa
The ${\cal B}^3_n$ are now elliptic fibrations over $F_{n/2}$ 
and still $K3$ fibrations over $P^1_{z_1}$.
Note that the unmodded 3-folds
$X^3_n$ and the modded
spaces ${\cal B}^3_n$ have still the same $K3$-fibres with Picard number 
$\rho$.
The Euler numbers of ${\cal B}^3_n$ can be computed from the Euler numbers
of $X^3_n$ from the ramified covering as
\beqa \chi(X^3_n)=2\chi ({\cal B}^3_n)-2\cdot 24.\label{eulercalb}
\eeqa
Using $\chi(X^3_n)=2(1+\rho-h^{(2,1)}(X^3_n))$ and
$\chi({\cal B}^3_n)=2+2(1+\rho-h^{(2,1)}({\cal B}^3_n))$ we
derive the followong relation between $h^{(2,1)}(X^3_n)$ and $h^{(2,1)}({\cal
B}^3_n)$:
\beqa
h^{(2,1)}({\cal B}^3_n)={1\over 2}(\rho-2+h^{(2,1)}(X^3_n)-19).
\label{hrelation}
\eeqa

Second, $X^4_n$ are of course no more products ${\cal B}^3_n\times T^2$
but the torus $T^2_{z_4}$ now is a second elliptic fibre which varies
over $P^1_{z_1}$. More precisely, this elliptic fibration describes just
the emergence of
the del Pezzo  surface $dP_9$, which is given in Weierstrass form as
\beqa
dP_9:\quad y^2=x^3+f_4(z_1)x+g_6(z_1).\label{weierdp}
\eeqa

Therefore the spaces $X^4_n$ have the form
of being the following fibre products:
\beqa
X^4_n=dP_9\times_{P^1_{z_1}}{\cal B}^3_n.\label{fibrep}
\eeqa
All $X^4_n$ are $K3$ fibrations
over the mentioned $dP_9$ surface.
The Euler numbers of all $X^4_n$'s are given by the value
\beqa
\chi=12\cdot 24=288.\label{euler}
\eeqa 
The corresponding (complex) three-dimensional IIB base manifolds $B^3_n$
have the following fibre product structure 
\beqa
B^3_n=dP_9\times_{P^1_{z_1}}F_{n/2}.\label{fibrebase}
\eeqa
For the case already studied in \cite{CL}  with $n=0$, $B^3_0$
is just the product space $dP_9\times P^1_{z_2}$.

The fibration structure of $X^4_n$ provides all necessary
informations to compute the Hodge numbers of $X^4_n$ from the number
of complex deformations of ${\cal B}^3_n$, which we call $N_{{\cal B}^3_n}$.
These can be calculated from eq.(\ref{weierb}) and are summarized in table
(\ref{tab4}).
Note that in the cases $n> 2$ we have to subtract in the $N=2$ setup
$7+n-1=6+n$ reparametrizations, whereas in the $N=1$ setup only $6+n/2$
(for $n=0,2$ we have to subract 7 reparamerizations both for $N=1$ and $N=2$).

Knowing that the number of complex deformations of $dP_9$ is eight, as it can
be easily be read off from eq.(\ref{weierdp}), we
obtain for the number of complex structure deformations of $X^4_n$ the
following result:
\beqa
h^{(3,1)}(X^4_n)=8+3+N_{{\cal B}^3_n}=11+N_{{\cal B}^3_n}.\label{h31}
\eeqa
Next compute the number of K\"ahler parameters, $h^{(1,1)}(X^4_n)$, of $X^4_n$.
Since $h^{(1,1)}(dP_9)=10$ we  obtain the formula
\beqa
h^{(1,1)}(X^4_n)=10+\rho ,\label{h11}
\eeqa
where $\rho$ is the Picard number of the $K3$ fibre of $X^4_n$.

Finally for the computation of $h^{(2,1)}(X^4_n)$ of $X^4_n$ we can use the
condition [\ref{sethi}] of tadpole cancellation, which tells us that
$h^{(1,1)}(X^4_n)-h^{(2,1)}(X^4_n)+h^{(3,1)}(X^4_n)={\chi\over 6}-8$.
Hence we get for $h^{(2,1)}(X^4_n)$
\beqa
h^{(2,1)}(X^4_n)=\rho+N_{{\cal B}^3_n}-19.\label{h21}
\eeqa
Using eqs.(\ref{h31},\ref{h11},\ref{h21}) we have summarized the 
spectrum of Hodge numbers
of $X^4_n$ in table (\ref{tab4}).
The computation of these 4-fold Hodge numbers, which was based on the counting
of complex deformations of the Weierstrass form eq.(\ref{weierb}), can be
checked in a rather independent way, by noting that $h^{(2,1)}(X^4_n)=
h^{(2,1)}({\cal B}^3_n)$,
since $N_{{\cal B}^3_n}=(19-\rho)+h^{(2,1)}({\cal B}^3_n)$. 
Then, using eq.(\ref{hrelation}) and
table (\ref{tab1}) the Hogde numbers $h^{(2,1)}(X^4_n)$ in table (\ref{tab4})
are immediately verified.

Let us compute the spectrum for our chain of models using 
eqs. (\ref{spectrumv},\ref{spectrumc}). 
The Hodge numbers of
$B^3_n$ eq.(\ref{fibrebase}) are universally given as
$h^{(1,1)}(B^3_n)=11$, $h^{(2,1)}(B^3_n)=0$.
Thus we obtain using eq.(\ref{h11}) that
\beqa
r(V)=h^{(1,1)}(X^3_n)-12=\rho -2;\label{v}
\eeqa
observe that the rank of the $N=1$ four-dimensional gauge groups
agrees with the rank of the six-dimensional gauge groups of the corresponding
$N=2$ parent models (see eq.(\ref{novector1})).\footnote{The Abelian
vector fields which arise in the $N=2$ situation from the $T^2$ compactification 
from six to four dimensions do not appear in the modded $N=1$ spectrum -- see
the discussion in the next chapter.}
Second, using eqs.(\ref{h31},\ref{h21}) we derive that
\beqa
C=38-r(V)+2h^{(2,1)}(X^4_n)=2+\rho+2N_{{\cal B}^3_n}.\label{c}
\eeqa
Using eq.(\ref{hrelation}), $C$ can be expressed by the number of 
hypermultiplets of the $N=2$ parent models as follows:
\beqa 
C=38+H-20.\label{ch}
\eeqa
This relation will become clear when considering in the
next chapter the dual heterotic models.
The explicit results for $N_{{\cal B}^3_n}$, $h^{(1,1)}(X^4_n)$, 
$h^{(2,1)}(X^4_n)$, $h^{(3,1)}(X^4_n)$, 
$r(V)$ and $C$ are contained in the
following table:

\beqa
%\begin{center}
\begin{tabular}{|c||c|c|c|c|c|c|} \hline
  $n$     & $N_{{\cal B}^3_n}$ & $h^{(1,1)}(X^4_n)$ & 
  $h^{(2,1)}(X^4_n)$ & $h^{(3,1)}(X^4_n)$ &
   $r(V)$  & $C$ \\ \hline\hline
0  & 129 & 12 & 112  & 140 & 0 & 262   \\ \hline
2 & 129 & 12  & 112  & 140 & 0 & 262   \\ \hline
4 &  141 & 16 & 128 & 152 & 4& 290 \\ \hline
6 & 165 & 18 & 154  & 176 & 6 & 340 \\ \hline
12 & 249 & 20 &  240 &  260 & 8 & 510 \\ \hline
\end{tabular}
%\end{center}
\label{tab4}
\eeqa

The equations (\ref{v}) and (\ref{c}) count the numbers of Abelian vector
fields and the number of neutral chiral moduli fields
of the four-dimensional $F$-theory compactification. 
%additional
%massless fields in general arise by adding various branes which may wrap
%around some of the non-trivial cycles of $X^4_n$. In fact, the existence
%of the branes is required by the cancellation of an $F$-theory anomaly
%which takes the value of $-\chi /24$. This anomaly can be either cancelled
%by $\chi /24$ 3-branes which fill the four-dimensional space time.
%Alternatively the anomaly can be also cancelled by non-trivial
%background gauge fields on the $F$-theory 7-branes.
Let us now discuss the emergence of $N=1$ non-Abelian gauge groups 
together with their matter contents. Namely, 
non-Abelian gauge groups arise by constructing  7-branes over which the
elliptic fibration has an ADE singularity
\cite{KatzV}. 
Specifically,
one has to consider
a (complex) two-dimensional space, which is a codimension one subspace of
the type IIB base $B^3$, over which the elliptic fibration has a singularity.
(In order to avoid adjoint matter the space $S$ must satisfy $h^{(2,0)}(S)=
h^{(1,0)}(S)=0$.)
The world volume of the 7-branes is then given by $R^4\times S$; if
$n$ parallel 7-branes coincide one gets for example an $SU(n)$ gauge
symmetry, i.e. the elliptic fibration acquires an $A_{n-1}$ singularity.
$N_F$ chiral massless
matter fields in the fundamental representation of the non-Abelian gauge
group can be geometrically engineered by bringing $N_f$ 3-branes
near the 7-branes, i.e close to $S$ \cite{BJPSV}.
The Higgs branches of these gauge theories should then be identified with
the moduli spaces of the gauge instantons on $S$. 
%So the pure Higgs branch
%corresponds to a situation where all 3-branes are replaced by the instantons,
%and the anomaly is entirely cancelled by  $\chi/24$ non-Abelian gauge
%instantons.
%The mixed branches are those with both 3-branes and instantons.
%We must emphasize that the formulas eqs.(\ref{v},\ref{c}) apply for the
%generic situation in moduli space, where the gauge group is Higgsed
%as much as possible.
%As we will show for our class of models this variety of Higgs branches 
%corresponds to the branches in the moduli spaces of bundles on the
%Calabi-Yau three-folds in the dual heterotic models.

%To be specific, since in our class of models the Euler number of all
%considered four-folds $X^4_n$ is $\chi=12\cdot 24=288$, the anomaly
%can be cancelled by 12 3-branes which fill space time.
In our class of models, 
the space $S$ is just given
by the
$dP_9$ surface which is the base of the $K3$ fibration of $X^4_n$.
The singularities of the elliptic four-fold fibrations are given by the
singularities of the Weierstrass curve for ${\cal B}^3_n$, given in 
eq.(\ref{weierb}). So the non-Abelian gauge groups arise at the
degeneration loci of eq.(\ref{weierb}). However with this observation
we are in the same situation as in the $N=2$ parent models, since
the singularities of the modded elliptic curve ${\cal B}^3_n$ precisely agree
with the singularities of the elliptic 3-folds $X^3_n$ in eq.(\ref{weier}).
In other words, the non-Abelian gauge groups in the $N=2$ and $N=1$
models are identical. This observation can be explained from the fact
that the gauge group enhancement already occurs in eight dimensions
at the  degeneration loci of the elliptic $K3$ surfaces as $F$-theory
backgrounds. However the underlying eight-dimensional $K3$ singularities
are not affected by the $Z_2$ operation on
the coordinate $z_1$, but are the same in eqs.(\ref{weier}) and (\ref{weierb}).

In the following section about the
heterotic dual models we will explicitly determine the non-Abelian
gauge groups and the possible Higgsing chains. We will show
that after maximal Higgsing of the gauge groups the dimensions
of the instanton moduli spaces, being identical the the dimensions 
of the Higgsing moduli spaces, precisely agree with $2h^{(2,1)}(X^4_n)-\rho+2$
on the $F$-theory side; 
in addition
we also verify that the ranks of the  unbroken gauge groups after
the complete Higgsing precisely match the ranks of the $F$-theory 
gauge groups, as given by $r(V)$ in table eq.(\ref{tab4}).

\vskip0.5cm

\section{Heterotic String on the $CY^{19,19}$}

\subsection{The $N=2$ models: the heterotic string on $K3(\times T^2)$}

In this section we will construct the  heterotic string compactifications
dual to $X^4_n$
with $N=1$ supersymmetry by $Z_2$ modding of $N=2$ heterotic string 
compactications which are the duals of the $F$-theory models on $X^3_n\times 
T^2$. 
Heterotic string models with $N=2$ supersymmetry in four 
dimensions are obtained
by compactification on $K3\times T^2$ plus the specification of
an $E_8\times E_8$ gauge bundle over $K3$.
So in the heterotic context, we have to specify how the $Z_2$ modding 
acts both on the compactification space $K3\times T^2$ as well
as on the heterotic gauge bundle.

The $N=2$ heterotic models,  that are dual to the 
$F$-theory compactifications
on $X^3_n\times T^2$, are characterized by turning on $(n_1,n_2)=
(12+n,12-n)$ ($n\geq 0$)
instantons of the heterotic gauge group $E_8^{(I)}\times E_8^{(II)}$ \cite{MV}.
Let us recall briefly the resulting spectrum, first in six 
dimensions on $K3$.
For this class of models there is one tensor multiplet which contains the
heterotic dilaton field. Next consider the massless vector and 
hypermultiplets
in six dimensions.
While the first $E_8^{(I)}$ is generically completely broken by the
gauge instantons, the second $E_8^{(II)}$ 
is only completely broken for the cases
$n=0,1,2$; for bigger values of $n$ there is a terminating gauge
group $G^{(II)}$ of rank $r(V)$ which cannot be further broken.
The quaternionic dimensions of the instanton moduli space of $n$
instantons of a gauge group $H$, living on  $K3$,
is in general given by
\beqa
\dim_Q({\cal M}_{\rm inst}(H,k))=c_2(H)n-\dim H,\label{n2inst}
\eeqa
where $c_2(H)$ is the dual Coxeter number of $H$.
Then, in the examples we are discussion, we derive the following formula
for the quaternionic dimension of the instanton moduli space:
\beqa
\dim_Q{\cal M}_{\rm inst}(E_8^{(I)}\times H^{(II)},n))=
112+30n+(12-n)c_2(H^{(II)})-\dim H^{(II)};
\label{hetinst}
\eeqa
here 
(for $n\neq 12$) $H^{(II)}$ is the 
commutant of the unbroken gauge group $G^{(II)}$ in 
$E_8^{(II)}$. Specifically, the following gauge groups $G^{(II)}$ and
dimensions of instanton moduli spaces are derived:

\beqa
%\begin{center}
\begin{tabular}{|c||c|c|} \hline
  $n$     & $G^{(II)}$  & $\dim_Q{\cal M}_{\rm inst}$ \\ \hline\hline
0  &   1 & 224   \\ \hline
2 &   1 & 224   \\ \hline
4 &    $SO(8)$& 252 \\ \hline
6 & $E_6$ & 302 \\ \hline
12 & $E_8$ & 472 \\ \hline
\end{tabular}
%\end{center}
\label{tab5}
\eeqa
 
The number of massless gauge singlet hypermultiplets is then simply given by
\beqa
H=20+\dim_Q{\cal M}_{\rm inst},\label{heth}
\eeqa
where the 20 corresponds to the complex deformations of $K3$.
It is well known that
comparing the spectra of $F$-theory on the 3-folds $X^3_n$
(see eqs.(\ref{nohyper},\ref{novector1}) and table (\ref{tab1})) with
the spectra of the heterotic string on $K3$ with instanton numbers $(12+n,12-n)$
(see eq.(\ref{heth}) and table (\ref{tab5})) one finds perfect
agreement.
Note that on the heterotic side there is an perturbative gauge symmetry
enhancement at special loci in the hypermultiplet moduli spaces.
Specifically, by embedding the $SU(2)$ holonomy group of
$K3$, namely the $SU(2)$ bundles with instanton numbers
$(12+n,12-n)$ in $E_8^{(I)}\times E_8^{(II)}$, the six-dimensional gauge group
is broken to $E_7^{(I)}\times E_7^{(II)}$ (or $E_7^{(I)}\times 
E_8^{(II)}$ for $n=12$); 
in addition one gets charged hyper multiplet fields, which can be used to
Higgs the gauge group via several intermediate gauge groups down to
the terminating groups. 
The dimensions of the Higgs moduli space,
i.e. the number of gauge neutral hypermultiplets, agrees with the dimensions
of the instanton moduli spaces eq.(\ref{hetinst}).

\subsection{The $N=1$ models: the heterotic string on $Z=(K3\times
T^2)/Z_2$}

Now let us construct the four-dimensional heterotic compactifications
with $N=1$ supersymmetry, which are dual to $F$-theory on $X^4_n$,
by $Z_2$ modding of the heterotic string compactications on $K3\times T^2$.
In the first step we discuss the $Z_2$ modding of 
the compactication space $K3\times T^2$ which
results in a particular Calabi-Yau 3-fold $Z$:\footnote{At the orbifold point
of $K3$ one can construct $Z$ as $T^6/(Z_2\times Z_2)$, where one of the
$Z_2$'s acts freely, see e.g. \cite{CL}.}
\beqa
Z={(K3\times T^2)\over Z_2}.\label{hetmodding}
\eeqa
Specifically, the $Z_2$-modding reduces $K3$  to the del Pezzo surface $dP_9$. 
This corresponds to having on K3 a Nikulin involution of type
(10,8,0) with two fixed elliptic fibers 
in the K3 leading to
\beqa
\begin{array}{ccc}K3&\rightarrow &dP_9\\\downarrow & &\downarrow\\P^1_y&
\rightarrow &P^1_{\tilde{y}}\end{array}
\eeqa
induced from the quadratic base map $y\rightarrow \tilde{y}:=y^2$ with the 
two branch points $0$ and $\infty$ (being the identity along the fibers).
In the Weierstrass representation of $K3$
\beqa
K3:\quad y^2=x^3-f_8(z)x-g_{12}(z),\label{weierk3}
\eeqa
the mentioned quadratic redefinition translates to the representation 
\beqa
dP_9:\quad y^2=x^3-f_4(z)x-g_6(z)\label{weierdpa}
\eeqa
of $dP_9$ (showing again the $8=5+7-3-1$ deformations).
So the $Z_2$ reduction of $K3$ to the
non Ricci-flat $dP_9$ corresponds to the
reduction of the $X^3_n$ to the non Calabi-Yau space ${\cal B}^3_n$
(cf. eqs.(\ref{weier} and \ref{weierb})).
Representing $K3$ as a complete intersection in the product of projective spaces
as $K3={\scriptsize \left[\begin{array}{c|c}P^2&3\\P^1&2\end{array}\right]}$,
the $Z_2$ modding reduces the degree in the $P^1$ variable by half; hence the
$dP_9$ can be represented as
$dP_9={\scriptsize \left[\begin{array}{c|c}P^2_x&3\\P^1_y&1\end{array}\right]}$. 
This makes 
visible on the one hand its elliptic fibration over $P^1$
via the projection onto the second factor; on the other hand the defining
equation $C(x_0,x_1,x_2)y_0+C^\prime (x_0,x_1,x_2)y_1=0$ shows that the 
projection onto the first factor exhibits $dP_9$ as being a $P^2_x$ blown up in
9 points (of $C\cap C^\prime$), so having as nontrivial hodge number (besides
$b_0,b_4$) only $h^{1,1}=1+9$. 
Furthermore the $dP_9$ has 8 complex structure moduli: they can be seen
as the parameter input in the construction of 
blowing up the plane in the 9 intersection points of two cubics
(the ninth of which is then always already determined as they sum up to
zero in the addition law on the elliptic curve;
so one ends up with $8\times 2-8$ parameters).

As in the dual $F$-theory description a second $dP_9$ emerges by fibering
the $T^2$ in eq.(\ref{hetmodding}) over the $P^1$ base of $dP_9$.
So in analogy to eq.(\ref{fibrep}) the heterotic Calabi-Yau 3-fold
$X^3_{\rm het}$, which is elliptically
fibred over $dP_9$, has the following fibre product structure
\beqa
Z=dP_9\times_{P^1} dP_9.\label{hetfibrep}
\eeqa
The number of K\"ahler  deformations of $Z$ is given by the sum of
the deformations of the two $dP_9$'s minus one of the common $P^1$ base,
i.e. $h^{(1,1)}(Z)=19$. Similarly we obtain
$h^{(2,1)}(Z)=8+8+3=19$.
This Calabi-Yau 3-fold is in fact well known being one of the Voisin-Borcea
Calabi-Yau spaces. It can be obtained from $K3\times T^2$ 
by the Voisin-Borcea involution, which consists in the `del Pezzo' involution
(type (10,8,0) in Nikulins classification) with two fixed elliptic fibers 
in the K3 combined with the 
usual ``-"-involution with four fixed points in the $T^2$. 
Writing $K3\times T^2$ as $K3\times T^2={\tiny 
\left[\begin{array}{c|cc}P^2&3&0\\P^1&0&2\\P^2&0&3\end{array}\right]}$
the Voisin-Borcea involution changes this to
$Z={\tiny 
\left[\begin{array}{c|cc}P^2&3&0\\P^1&1&1\\P^2&0&3\end{array}\right]}=
dP\times _{P^1}dP$.
%So here the symmetric degree one entries in the 
%$P^1$ variables have a seemingly different origin: one by {\it `reduction'} 
%(from 
%two) and one by {\it `emergence'} (from zero).
Observe that the base of the elliptic fibration of $Z$ is given
by the $dP_9$ surface which {\it `emerges'} 
(from the trivial elliptic fibration) after the $Z_2$ modding.

After having described the $Z_2$ modding of $K3\times T^2$ we will now
discuss how this operation acts on the heterotic gauge bundle.
Recall that in the $N=2$ heterotic models on $K3\times T^2$ the heterotic
gauge group $E_8^{(I)}\times E_8^{(II)}$ is living on the four-dimensional
space $K3$. We will now consider a $N=1$ situation where after the $Z_2$
modding the heterotic gauge group still lives on a four manifold, namely on
the del Pezzo surface $dP_9$, which arises from the $Z_2$ modding of
the $K3$ surface.
Then the complex dimension of the instanton moduli space of $k$
gauge instantons of a gauge group $H$, which lives on $dP$, is given by
(see for example \cite{VZ})
\beqa
\dim_C {\cal M}_{\rm inst}(H,k)=2c_2(H)k-\dim H.\label{complexdim}
\eeqa

The action of the $Z_2$ modding on the gauge bundle is now defined 
in such a way that
the gauge instanton numbers are reduced by half (think of the limit case of
pointlike instantons):
\beqa
k_{1,2}={n_{1,2}\over 2}.\label{nkhalfe}
\eeqa
So the total number of gauge instantons in $E_8^{(I)}
\times E_8^{(II)}$ will be reduced by two, i.e. $k_1+k_2=12$ and we are
considering $(k_1,k_2)=
(6+{n\over 2},6-{n\over 2})$ instantons in $E_8^{(I)}\times E_8^{(II)}$.
The reduction of the total instanton number by half from 24 to 12
can be explained from the observation that on the $F$-theory side
the tad-pole anomaly can canceled either by $\chi/24=12$
3-branes  or by 12
gauge instantons of the  gauge group $H$, which lives over the
four manifold $S=dP_9$.
%Therefore one needs either 12 5-branes in the heterotic dual models
%\cite{FMW} or the background of 12 gauge instantons. 
So, with $k_1+k_2=12$ and using eq.(\ref{complexdim})
we can compute the complex dimensions of the instanton moduli space for the
gauge group $E_8^{(I)}\times H^{(II)}$, where again $H^{(II)}$ is the commutant
of the gauge group $G^{(II)}$ which cannot be further broken by the
instantons:
\beqa
\dim_C{\cal M}_{\rm inst}(E_8^{(I)}\times H^{(II)},n))=
112+30n+(12-n)c_2(H^{(II)})-\dim H^{(II)}.\label{hetinsta}
\eeqa
 This result precisely agrees with the quaternionic dimensions of the
instanton moduli space, eq.(\ref{hetinst}), in the unmodded $N=2$ models.
So we see that we obtain 
as gauge bundle deformation parameters of the heterotic string on 
$Z$
the same number of massless, gauge neutral $N=1$
{\it chiral} multiplets as the number of massless $N=2$ {\it hyper} multiplets
of the heterotic string on $K3$.
This means that the $Z_2$ modding  keeps 
in the massless sector just one of the two chiral fields
in each $N=2$ hyper multiplet.
These chiral multiplets describe the Higgs phase of the $N=1$ heterotic
string compactifications.

The gauge fields in $N=1$ heterotic string compactifications on $Z$
are just given by those gauge fields which arise from the compactification
of the heterotic string on $K3$ to six dimensions; therefore they are
invariant under the $Z_2$ modding. However the complex scalar fields
of the corresponding $N=2$ vector multiplets in four dimensions
do not survive the $Z_2$ modding. Therefore there is no Coulomb phase
in the $N=1$ models in contrast to the  $N=2$ parent compactifications.
Also observe that the two vector fields, commonly denoted by $T$ and $U$,
which arise from the compactification from six to four dimensions on $T^2$
disappear from the massless spectrum
after the modding. This is expected since the
Calabi-Yau space has no isometries which can lead to massless gauge
bosons. Finally, the $N=2$ dilaton vector multiplet $S$ is reduced
to a chiral multiplet in the $N=1$ context.

These relations between the spectra of the $N=1$ and $N=2$ models can
be understood from the observation that the considered ${\bf Z}_2$ modding
corresponds to a spontaneous breaking of $N=2$ to $N=1$ spacetime
supersymmetry \cite{KK}.

In summary, turning on $(6+{n\over 2},6-{n\over 2})$
gauge instantons of $E_8^{(I)}\times E_8^{(II)}$
in our class of $N=1$ heterotic string compactifications on
$Z$, the unbroken gauge groups $G^{II}$ as well as the number
of remaining massless chiral fields 
(not counting the geometric moduli from $Z$, see next paragraph) 
agree with the unbroken gauge groups
and the number of massless hyper multiplets (again without
the 20 moduli from $K3$) in the  heterotic 
models on $K3$ 
with $(12+n,12-n)$ gauge instantons.
The specific gauge groups and the numbers of chiral fields are 
already summarized in table (\ref{tab5}).

Now comparing with the  $F$-theory spectra we first observe
that the ranks of the gauge groups after
maximally possible Higgsing perfectly match in the two dual
descriptions (see tables (\ref{tab4}) and (\ref{tab5})).

Next compare the number  of chiral $N=1$  
moduli fields in the heterotic/F-theory
dual pairs. 
First, looking at the Hodge numbers of the dual $F$-theory fourfolds
$X^4_n$, as given in table (\ref{tab4}) we recognize that
\beqa 
2h^{(2,1)}(X^4_n)-(\rho -2)=\dim_C {\cal M}_{\rm inst}.\label{agree}
\eeqa

Let us give an argument for this independent of the case by case
calculation. 
Namely, using $C^F=38+(2h^{(2,1)}(X^4_n)-(\rho-2))=H+18$ (cfr. eqs.(\ref{c}) and
(\ref{ch})) and $H=20+\dim_Q{\cal M}_{\rm inst}=20+\dim_C{\cal M}_{\rm inst}$,
%$e_{CY^3}=2\cdot e_{{\cal B}}-2\cdot 24$ by the ramified covering
%and $e_{CY^3}=2(1+\rho -h^{2,1}(CY^3))$,
%$e_{{\cal B}}=2+2(1+\rho -h^{2,1}({\cal B}))$ one gets the result
%from the number of hypermultiplets 
%$\sharp H=h^{2,1}(CY^3)+1=20+\dim_Q {\cal M}_{inst}$ 
%and using that 
%$h^{2,1}(X^4)=h^{2,1}({\cal B})$ as
%$N_{{\cal B}^3_n}=(19-\rho)+h^{2,1}({\cal B})\leftrightarrow
%h^{3,1}=8+3+N_{{\cal B}^3_n}=30-\rho +h^{2,1}({\cal B})\leftrightarrow
%h^{2,1}=h^{1,1}+h^{3,1}-40=h^{2,1}({\cal B})$.
%
%SHOW $N_{{\cal B}^3_n}=(19-\rho)+h^{2,1}({\cal B})$.
%
one gets 
\beqa
C_F=38+(2h^{(2,1)}(X^4_n)-(\rho -2))=38+\dim_C {\cal M}_{\rm inst}
\eeqa

On the 
heterotic side 
the total number $C$ of chiral moduli fields is 
given by the dimension of the gauge
instanton moduli space plus the number of geometrical moduli 
$h^{(1,1)}(Z)+h^{(2,1)}(Z)$
from the
underlying Calabi-Yau space $Z$, 
which is 38 for our class of models, i.e.
\beqa
C_{\rm het}=38+\dim_C {\cal M}_{\rm inst}.\label{chet}
\eeqa

So we have shown that the massless spectra
of Abelian vector multiplets and of the gauge singlet chiral plus
antichiral fields agree for all considered dual pairs.
The next step in the verification of the $N=1$ string-string duality
after the comparison of the massless states
is to show that the interactions, i.e. the $N=1$ effective action,
agree. In particular, the construction of the superpotentials
is important to find out the ground states of these theories.
This was already done \cite{DGW,CL} for one particular model 
(the model with $k=9$, $n=0$), where on the heterotic
side the superpotential was entirely generated by world sheet instantons.
It would be interesting to see, whether also space time instantons
would contribute to the heterotic superpotential in some other models 
and whether supersymmetry can be broken by the
superpotential. In addition, it would be also interesting to compare
the holomorphic gauge kinetic functions in $N=1$ dual string pairs, in
particular in those models which are obtained from $N=2$ dual pairs
by $Z_2$ moddings respectively by spontaneous supersymmetry
breaking from $N=2$ to $N=1$.

\subsection{Non-Abelian gauge groups and Higgsing chains}

For the computation of $C$ and $r(V)$ we have considered  a generic
point in the moduli space where the gauge group is broken as far as possible
to the group $G^{II}$
by the vacuum expectation values of the chiral fields.
In this section we now want to determine the non-Abelian gauge groups
plus their matter content which arise in special loci of the
moduli space. 
Since the $N=1$ gauge bundle is identical to the $N=2$ bundle, which is given
by $SU(2)\times SU(2)$, the maximally unbroken gauge group is
for $k_1,k_2\geq 3$ (i.e. $n\leq 6$)
given by $E_7^{(I)}\times E_7^{(II)}$;
the $N=1$ chiral representations follow immediately from the
$N=2$ hypermultiplet representations and transform as

\beqa
E_7\times E_7:\quad (k_1-2) ({\bf 56},{\bf 1})  
 +
(k_2-2) ({\bf 1},{\bf 56})
+ (4(k_1+k_2)-6)({\bf 1},{\bf 1}).
\label{e7}
\eeqa
Higgsing $E_7^{(I)}\times E_7^{(II)}$ to $E_6^{(I)}\times E_6^{(II)}$
 one is left with  
chiral matter fields in the following representations of the
gauge group $E_6^{(I)}\times E_6^{(II)}$:
\beqa
E_6\times E_6:\quad (k_1-3)\lbrack ({\bf 27},{\bf 1}) + ({\bf\overline{27}},
{\bf 1})\rbrack &+&
(k_2-3)\lbrack ({\bf 1},{\bf 27})
+({\bf 1},{\bf\overline{27}})\rbrack \nonumber\\ 
+ (6(k_1+k_2)-16)({\bf 1},{\bf 1}).
\label{e6}
\eeqa
When $k_1=12$ ($n=12$) the gauge group is $E_6^{(I)}\times E_8^{(II)}$
with chiral matter fields
\beqa
E_6\times E_8:\quad 
9\lbrack ({\bf 27},{\bf 1}) + ({\bf\overline{27}},{\bf 1})\rbrack 
+ 64({\bf 1},{\bf 1}).\label{e6e8}
\eeqa
Since $k_1\geq 6$, the number of $({\bf 27},{\bf 1}) + 
({\bf\overline{27}},{\bf 1})$ is always big enough that the first
$E_6^{(I)}$ can be completely broken. On the other hand, only
for the cases $n=0,2$ the group $E_6^{(II)}$ can be completely Higgsed away
by giving vacuum expectation values to the fields
$({\bf 1},{\bf 27})
+({\bf 1},{\bf\overline{27}})$.
For $k_2=4$ ($n=4$), $E_6^{(II)}$ can be only Higgsed to the group 
$G^{(II)}=SO(8)$, and
for $k_2=3$ ($n=6$) 
there are no charged fields
with respect to $E_6^{(II)}$ such that the terminating gauge group is just 
$E_6^{(II)}$.
Clearly, the ranks of these
gauge groups are in agreement with the previous discussions, i.e.
with the results for $r(V)$; in addition,
assuming maximally possible Higgsing of both gauge group factors the complex
dimension
of the Higgs moduli space agrees with the dimensions of the instanton
moduli spaces as given in eq.(\ref{hetinsta}) and in table (\ref{tab5}).

Consider the Higgsing of, say, the first gauge group $E_6^{(I)}$.
Namely,
like in the
$N=2$ cases \cite{CF}, it 
can be Higgsed through the following chain of Non-Abelian
gauge groups:
\beqa
E_6\rightarrow SO(10)\rightarrow SU(5)\rightarrow SU(4)\rightarrow
SU(3)\rightarrow SU(2)\rightarrow SU(1).\label{chain}
\eeqa
In the following we list the spectra for all gauge groups within this chain:
\beqa
SO(10)& :&
\quad
(k_1-3)( {\bf 10} + {\bf\overline{10}})  +
(k_1-4)({\bf 16}+{\bf\overline{16}})+(8k_1-15){\bf 1}
,\label{so10}\\
%\eeqa
%\beqa
SU(5)& :&
\quad
(3k_1-10)( {\bf 5} + {\bf\overline{5}})  +
(k_1-5)({\bf 10}+{\bf\overline{10}})+(10k_1-24){\bf 1}
,\label{su5}\\
%\eeqa
%\beqa
SU(4)& :&
\quad
(4k_1-16)( {\bf 4} + {\bf\overline{4}})  +
(k_1-5)({\bf 6}+{\bf\overline{6}})+(16k_1-45){\bf 1}
,\label{su4}\\
%\eeqa
%\beqa
SU(3)& :&
\quad
(6k_1-27)( {\bf 3} + {\bf\overline{3}})  +(24k_1-78){\bf 1}
,\label{su3}\\
%\eeqa
%\beqa
SU(2)& :&
\quad
(12k_1-56) {\bf 2} +  (36k_1-133){\bf 1}
,\label{su2}\\
SU(1)& : & \quad (60 k_1-248){\bf 1}.\label{su1}
\eeqa
In order to keep  contact with our previous discussion, we see that
the number of massless chiral fields at a generic point in the
moduli space, i.e. for complete Higgsing down to $SU(1)$, is given
by $60 k_1-248 +2k_2c_2(H^{(II)})-\dim H^{(II)}$ which precisely
agrees with $\dim_C{\cal M}_{\rm inst}$.

\vskip0.5cm

{\bf Acknowlegement}
We thank S. Kachru for discussions.
The work is supported by NAFOEG, the Deutsche Forschungsgemeinschaft (DFG)
and by the European Commission TMR programme ERBFMRX-CT96-0045.

\vskip0.5cm

%%%%%%%%%%%%%%%%%%%%% Appendix A  %%%%%%%%%%%%%%%%%%%%%%%%%%%%%%%%%
%\appendix

%\begin {appendix}
%{\Large\bf Appendix}\\

%\section{Appendix}

\setcounter{equation}{0}

%%%%%%%%%%%%%%%%%%%%%%%%%%%%%%%%%%%%%%%%%%%%%%%%%%

\end{document}